\documentclass[sigconf]{acmart}
\AtBeginDocument{%
  }

\setcopyright{acmlicensed}
\copyrightyear{2026}
\acmYear{2026}
\acmDOI{}
\acmConference[IEEE/ACM 7th International Workshop on Software Engineering Research \& Practices for the IoT (SERP4IoT)]{April 12--18,
  2026}{Rio de Janeiro, Brazil}

\usepackage[utf8]{inputenc}
\usepackage{xcolor}
\usepackage{enumitem}
\usepackage{hyperref}
\usepackage{color}
\usepackage{listings}
\usepackage[T1]{fontenc}
\usepackage{multirow}
\usepackage{colortbl}
\usepackage{url}
\usepackage{comment}
\usepackage{colortbl}
\usepackage{lipsum}
\usepackage{multicol}
\usepackage{amsmath}
\usepackage{graphicx}
\usepackage{algorithmic}
\usepackage{listings}
\usepackage{amsfonts}
\usepackage{tcolorbox}
\usepackage[linesnumbered, ruled, vlined]{algorithm2e}
\usepackage{longtable}
\usepackage{todonotes}
\usepackage{acronym}
\usepackage{lscape} 

\definecolor{lightgray}{rgb}{.9,.9,.9}
\definecolor{darkgray}{rgb}{.4,.4,.4}
\definecolor{darkgreen}{rgb}{0, 0.39, 0.00}
\definecolor{Gray}{gray}{0.7}
\definecolor{codegreen}{rgb}{0,0.6,0}
\definecolor{codegray}{rgb}{0.5,0.5,0.5}
\definecolor{codepurple}{rgb}{0.58,0,0.82}
\definecolor{backcolour}{rgb}{0.95,0.95,0.92}

\lstdefinestyle{mystyle}{
    backgroundcolor=\color{backcolour},   
    commentstyle=\color{codegreen},
    keywordstyle=\color{magenta},
    numberstyle=\tiny\color{codegray},
    stringstyle=\color{codepurple},
    basicstyle=\ttfamily\footnotesize,
    breakatwhitespace=false,         
    breaklines=true,                 
    captionpos=b,                    
    keepspaces=true,                 
    numbers=left,                    
    numbersep=5pt,                  
    showspaces=false,                
    showstringspaces=false,
    showtabs=false,                  
    tabsize=2
}

\acrodef{ECU}{Electronic Control Unit}
\acrodef{ML}{Machine Learning}
\acrodef{NN}{Neural Networks}
\acrodef{RT}{Real-Time}
\acrodef{ACC}{Adaptive Cruise Control }
\acrodef{CPS}{Cyber Physical System}
\acrodef{IDS}{Intrusion Detection System}
\acrodef{CAN}{Controller Area Network}
\acrodef{HMM}{Hodden Markov Model}
\acrodef{UAV}{Unmanned Aerial Vehicle}
\acrodef{EKF}{Extended Kalman Filter}
\acrodef{SITL}{Instrumented Software-in-the-Loop}
\acrodef{PID}{Proportional-Integral-Derivative}
\acrodef{GCS}{Ground Control Station}
\acrodef{HAL}{Hardware Abstraction Layer}
\acrodef{AHRS}{Attitude and Heading Reference System}
\acrodef{GPS}{Global Positioning System}
\acrodef{PWM}{Pulse Width Modulation}
\acrodef{IMU}{Inertial Measurement Unit}
\acrodef{ESC}{Electronic Speed Controller}
\acrodef{SDA}{Sensor Deprivation Attack}
\acrodef{EKF}{Extended Kalman Filter}
\acrodef{RE}{Reverse Engineering}
\acrodef{HAL}{Hardware Extraction Layer}
\acrodef{RCU}{Remote Controller Unit}
\acrodef{NED}{North-East-Down}
\acrodef{EKF3} {Enhanced Kalman Filter V3}
\acrodef{SITL} {software-in-the-loop}
\acrodef{RC}{Remote Control}

\begin{document}

\title{Reverse Engineering and Control-Aware Security Analysis of the ArduPilot UAV Framework}

\author{Yasaswini Konapalli$^{1}$, Lotfi Ben Othmane$^{1}$, Cihan Tunc$^{1}$, Feras Benchellal$^{1}$, Likhita Mudagere$^{1}$, \\
	\normalsize $^1$University of North Texas, Denton, TX, USA
}

\begin{abstract}
\ac{UAV} technologies are gaining high interest for many domains, which makes UAV security of utmost importance. 
ArduPilot is among the most widely used open-source autopilot \ac{UAV} frameworks; yet, many studies demonstrate the vulnerabilities affecting such systems. Vulnerabilities within its communication subsystems (including WiFi, telemetry, or GPS) expose critical entry points, and vulnerabilities in Ardupilot can affect the control procedure. In this paper, we reconstruct the software architecture and the control models implemented by ArduPilot and then examine how these control models could be misused to induce malicious behaviors while relying on legitimate inputs. 
\end{abstract}

\ccsdesc[10003022]{Security and privacy → Software and application security → Software reverse engineering}
  
\maketitle

\section{Introduction} 

Unmanned Aerial Vehicles (UAVs) are now integral to agriculture, environmental monitoring, aerial imaging, logistics, and defense due to their autonomy and adaptability, including in-flight mission/route updates~\cite{mohsan2023unmanned}. 
As a result, the overall drone market was estimated to be around USD \$27.4 billion in 2021 and is projected to reach USD \$58.4 billion by 2026~\cite{globenewswire_2023}. 
Nevertheless, the in-flight update flexibility also increases the UAV attack surface~\cite{tufekci2021vulnerability}. For example, in its default form, MAVLink (Micro Air Vehicle Link -- a lightweight messaging protocol for communicating with drones (and between onboard drone components)~\cite{koubaa2019micro}) lacks encryption and authentication, enabling adversaries to inject or replay commands, spoof telemetry, or mislead operators~\cite{tufekci2024enhancing}. 
Another major concern is ArduPilot-related risks~\cite{Ardupilot,10740195,MEKDAD2023109626} as ArduPilot is among the most widely used open-source autopilot stacks operating on over a million vehicles~\cite{Ardupilot.org} 
providing flight modes, sensor fusion, motor control, and communications~\cite{wang2021exploratory}. 

 \begin{figure}[htbp]
     \centering
     \includegraphics[width=0.75\linewidth]{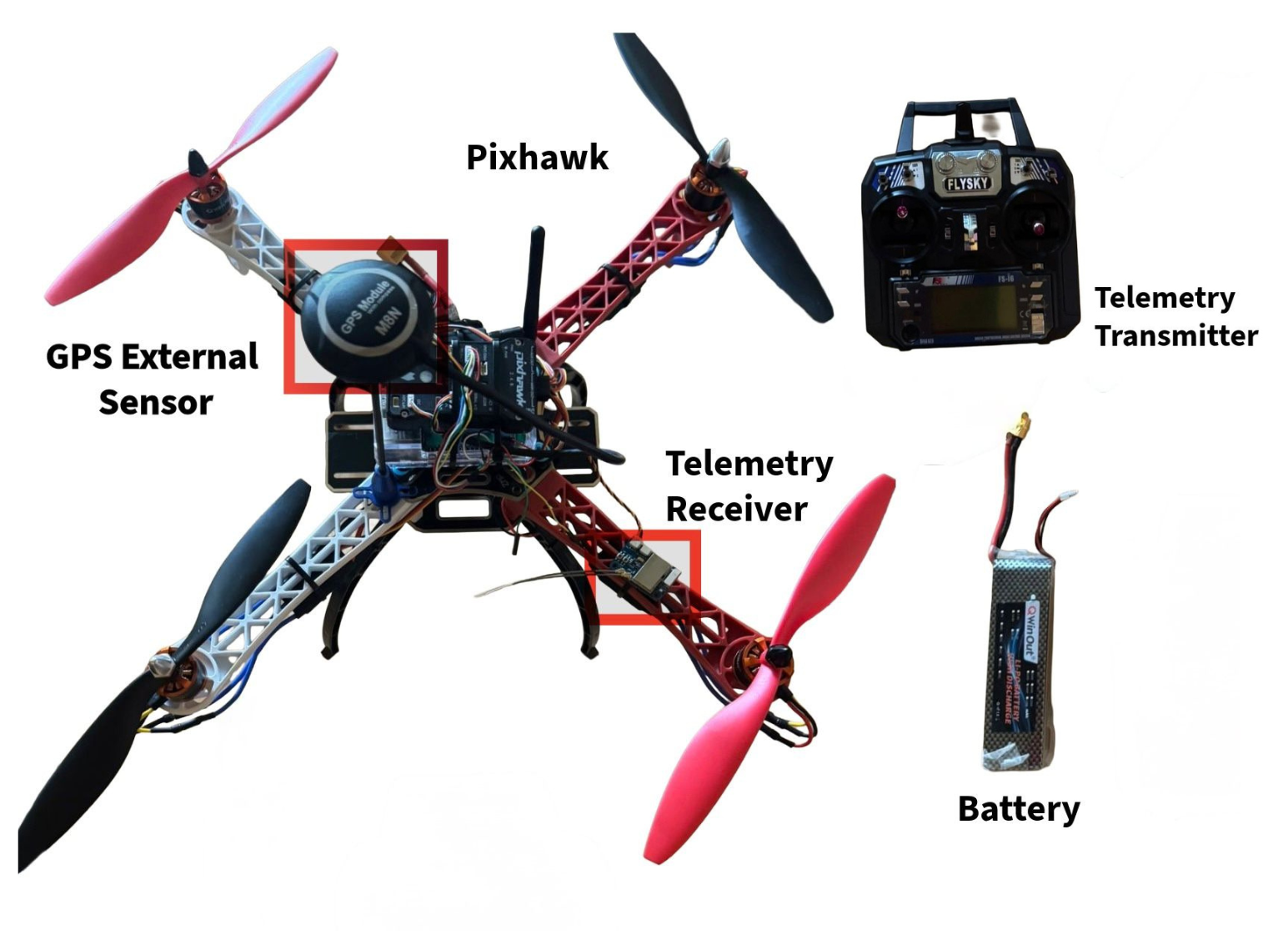}
          \caption{Experimental hardware setup showing Pixhawk~4 flight controller, GPS, telemetry radio, power distribution, and a ground control station.}
     \label{fig:Pixhawk}
     \vspace{-4mm}
 \end{figure}

Even though many existing works emphasize link-level protections(e.g.,~\cite{marty2014mavlink,kwon2018mavlink,erba2024sda,jeong2021muvids}), they do not address how ArduCopter firmware parses and executes redirection pathways internally or how legitimate-looking commands can induce unsafe control responses. UAV controllers typically assume commands and sensor inputs are trustworthy; hence, if these inputs are manipulated, the control loop may produce destabilizing actions despite being correct by design. 
Hence, this paper investigates the following question: 
\emph{Can legitimate control commands be used to mislead ArduPilot?} 
A positive answer to this question would shift the focus from traditional network-level security threats, such as spoofing or injection attacks, toward vulnerabilities within the control models and software implementation itself. This shift could promote the development of new techniques aimed at mitigating such internal security weaknesses.

We investigate mid-flight redirection attacks against ArduCopter (a subset of ArduPilot as a specific firmware for multicopters and helicopters) by combining reverse engineering with \ac{SITL} experimentation. We map the code paths that accept guided/mission-override commands, trace their propagation through flight-mode logic and controllers, and discuss injections that redirect a vehicle without firmware exploits. 
The main contributions of this paper are as follows:
\begin{enumerate}
\item Reverse-engineering of ArduPilot's internal control structure.
\item Reconstruction of cascaded PID control models for horizontal, vertical, attitude, and rate control.
\item Analysis of attack surfaces within the control models, including MAVLink command injection, parameters tampering, sensors spoofing, and exception-handling weaknesses.
\item Analysis of these contributions using a multirotor platform built around a Pixhawk~2.4.8 running ArduPilot (v3) built with the GNU ARM Embedded Toolchain and deployed via Mission Planner~1.3.83 with the integration of an \ac{IMU}, barometer, and magnetometer sensors to provide attitude, altitude, and heading estimation as well as an external GPS. We use a FlySky transmitter/receiver for manual control and for failsafe during flight tests. 
We validated our experiments using Software-in-the-Loop (SITL) first, and then replicated using hardware. 
\end{enumerate}

The remainder of this paper is organized as follows. Section~\ref{sec:relworks} reviews related work and outlines the position of this study. Section~\ref{sec:architecture} presents the reverse-engineering approach and provides sketches of the extracted architecture. Section~\ref{sec:implementedControllers} enumerates and describes the controllers implemented in Ardupilot. Finally, Section~\ref{sec:analysis} analyzes potential manipulations of Ardupilot through the implemented control models and exception-handling mechanisms.


\section{Related work}\label{sec:relworks}

\noindent{\bf MAVLink protocol vulnerabilities:} The MAVLink protocol has been exploited as an entry point to AV systems. 
Marty was among the first to assess the vulnerabilities of the MAVLink protocol used for UAV command and control~\cite{marty2014mavlink}. The study demonstrated that unencrypted and unauthenticated MAVLink messages can expose critical weaknesses in confidentiality, integrity, and availability. Using real-world testing with the ArduPilot Mega 2.5 and Mission Planner, this work showed that message injection, hijacking, and denial-of-service attacks could compromise UAVs in flight. 
Building on this, Kwon et al. conducted an empirical analysis of MAVLink vulnerabilities, confirming that a remote adversary can exploit unprotected packets to disable a mission or force a UAV to hover mid-operation using false packet injections and flooding attacks~\cite{kwon2018mavlink}. 

\noindent{\bf Sensor and control-loop manipulation:} The UAVs heavily depend on various sensors for operation success, including but not limited to GPS, Inertial Measure Unit (IMU) for attitude (roll, pitch, and yaw), velocity, and changes in altitude and gravitational forces, etc. To exploit sensor reconfiguration via message injection rather than continuous spoofing, Erba et al. introduced Sensor Deprivation Attacks (SDA)~\cite{erba2024sda}, selectively suspending IMU updates through malicious I²C (Inter-Integrated Circuit) messages, which allows attackers to stall the Kalman Filter and destabilize flight control loops without altering firmware or visible telemetry. 
This research expands the threat model to sensor-level interference, showing how attackers can indirectly redirect or crash a UAV by manipulating internal timing and control synchronization. 

\noindent{\bf Network-level detection and false injection defense:} 
Jeong et al. introduced MUVIDS, an intrusion detection system that targets MAVLink-enabled UVs~\cite{jeong2021muvids} by leveraging LSTM-based sequential models to identify false command injection patterns at the network level. 
This study validated that deep learning classifiers could effectively detect three types of MAVLink message injections across software-in-the-loop and hardware-in-the-loop environments. This work shifts from vulnerability discovery to behavioral anomaly detection in communication, highlighting that false MAVLink messages could still be injected when authentication or message signing is disabled, hence emphasizing the need for real-time protection mechanisms at the protocol level without altering onboard firmware. 

\noindent{\bf Software reliability:} Research on open-source autopilot systems (e.g., Pixhawk-based platforms and ArduPilot) highlights their widespread adoption but also exposes inherent reliability and security issues. The ArduPilot project was subjected to monthly vulnerability scans using Coverity until 2019, during which a total of 889 software defects were identified over four years~\cite{coverityardupilot}. Most of these defects were associated with illegal memory access, uninitialized variables, memory leaks, and data corruption. The regular scans were discontinued in 2019, leaving 431 unresolved defects.

Wang and his team conducted the first large-scale empirical study of autopilot software bugs, where they analyzed 569 real issues across Pixhawk and ArduPilot~\cite{wangpaper}. They identified eight UAV-specific bug categories: (1) \textit{Limit bugs} caused by incorrect parameter limits, such as allowing negative values; \textit{math bugs} misusing mathematical formulas for estimation or control, (3) \textit{inconsistency bugs} or hardware–software mismatches, (4) \textit{priority bugs}, (5) \textit{parameter bugs} misusing or missing a parameter, (6) \textit{hardware support bugs}, (7) correction bugs, and (8) initialization bugs. These bugs can cause flight instability, crashes, or undefined behavior, making software reliability a foundational security issue. The study emphasizes that discovering and addressing UAV bugs is not an easy task because standard comparisons are not efficient, which leaves developers to rely on logs, parameters, and trial and error.

The persistent occurrence of software vulnerabilities within the UAV control mechanisms raises a critical research question: \textit{What is the impact of exploiting such vulnerabilities on the behavior and safety of UAV systems? }While this study does not directly assess the effects of vulnerabilities, it instead aims to demonstrate how indirect command manipulation can influence the behavior and safety of UAV systems.

\begin{figure*}[tb]
        \centering
        \includegraphics[width=0.7\linewidth]{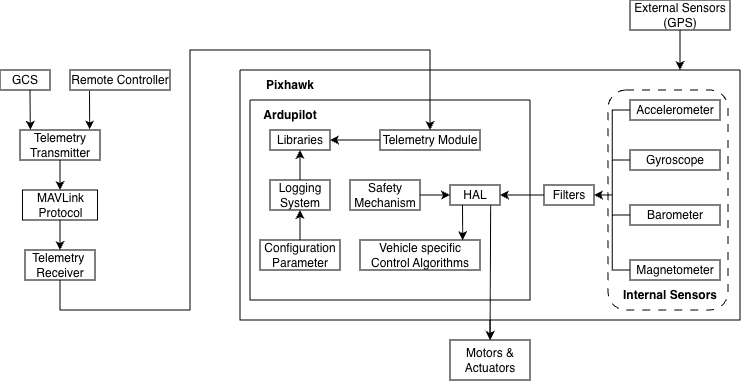}
        \caption{Ardupilot Architecture.}
        \label{fig:ardupilot_arch}
\end{figure*}

\section{ArduPilot Architecture}\label{sec:architecture}

Although documentation for ArduPilot (and subsequently ArduCopter) exists, the available information on its internal architecture and decision-making factors is not well-documented. Thus, through this process, we aim to extract/reconstruct the internal architecture of ArduPilot to reveal how user commands propagate through its control hierarchy down to actuator execution through \ac{RE}. 
For this purpose, we used manual static code inspection and 
(\ac{SITL}) experimentation, including call-graph tracing to map the data flow from MAVLink commands to motor actuation. Static analysis of the ArduCopter codebase identified the main entry points and class relationships (e.g., \texttt{Copter.cpp}), mode handlers (\texttt{ModeGuided}, \texttt{ModeAuto}, \texttt{ModeLoiter}), and controller modules (\texttt{AC\_PosControl}, \texttt{AC\_AttitudeControl}, \texttt{AP\_Motors}) and established how MAVLink commands are parsed and translated into position, velocity, and attitude targets.

ArduPilot employs a \textbf{scheduler} that periodically executes a deterministic sequence of control tasks, such as \ac{IMU} update, rate control, and motor output. Each task's frequency and dependency are explicitly defined, enabling analysis of loop timing and itter (the variation/deviation in task execution intervals). This structure allows repeatable timing verification and assessment of real-time performance consistency within the flight control stack.



Figure~\ref{fig:ardupilot_arch} illustrates the extracted overall architecture of ArduPilot. 
The communication between the \ac{GCS}/remote controller and the UAV is through a telemetry transmitter and receiver using MAVLink protocol. The telemetry receiver sends the messages to the Pixhawk hardware running ArduPilot, which uses various internal (e.g., accelerometer, gyroscope, barometer) and external sensors (\ac{GPS}) for operations~\cite{Pixhawk}. 
During system boot, the firmware loads configuration parameters, performs arming checks, and registers periodic tasks. 
The platform provides several core services: 
(a) \acf{HAL}: Hardware abstraction for sensors, actuators, timers, and serial/CAN/I2C interfaces. 
(b) Telemetry: MAVLink routing, mission and parameter protocols, heartbeat generation, and time synchronization. 
(c) Logging and Configuration: Structured DataFlash logging and runtime parameter management. 
(d) Safety: Link-loss failsafes, geofencing, watchdog monitoring, and flight termination mechanisms. 
Finally, ArduPilot implements a finite-state machine that governs flight behavior according to the selected mode (e.g., Manual, Guided, Auto, Loiter, or Return-to-Launch) by processing inputs from the \ac{GCS} and onboard scripts. The computed outputs (i.e., thrust and torques) are then converted by the \ac{HAL} mixer into \ac{PWM} or digital 
commands, ensuring platform portability across Pixhawk-compatible boards.


ArduCopter extends ArduPilot for multirotor vehicles, integrating flight-mode logic, guidance, cascaded PID control, and geometry-specific mixing. It adopts the \ac{NED} reference frame ($x = N$, $y = E$, $z = D$). Core control components, including sensor fusion via \ac{EKF3}~\cite{10.1115}, and \ac{PID}-based position, velocity, and attitude controllers, provide a consistent, drift-free state estimation for closed-loop stability. 
%
%
The \ac{EKF3} runs a two-step predict–update process. In the prediction step, inertial data propagate the system state shown in Eq.~\ref{eq:EKF3_pred}.

\begin{align}\label{eq:EKF3_pred}
\nonumber x_{t|t-1} = f(x_{t-1}, u_{t-1}, w_{t-1}), \\
P_{t|t-1} = F_t P_{t-1} F_t^T + Q_t
\end{align}
where $x$ is the state vector, $t$ is the step, $f$ is the function of the position velocity, $P$ the covariance,$F_t$ is the Jacobian of $f$ with respect to $x_{t-1}$ and $Q_t$ the process noise covariance. Upon receiving new measurements $z_t$ (from GPS, barometer, or magnetometer), the correction step is applied in Eq.~\ref{eq:EKF3_correction}:
\begin{align}\label{eq:EKF3_correction}
\nonumber K_t = P_{t|t-1} H_t^T (H_t P_{t|t-1} H_t^T + R_t)^{-1}, \\
x_t = x_{t|t-1} + K_t (z_t - h(x_{t|t-1}))
\end{align}
where $K_t$ is the Kalman gain, $H_t$ the measurement model Jacobian, and $R_t$ the measurement noise covariance. 
This recursive estimation ensures robustness against sensor noise and delays. EKF outputs are fed into the cascaded controllers for position, velocity, and attitude, enabling stable flight even under disturbances or manipulated MAVLink.

\section{ArduCopter Control Models}\label{sec:implementedControllers}


\subsection{Overview of Cascaded Control}

\begin{figure}
    \centering
    \includegraphics[width=0.99\linewidth]{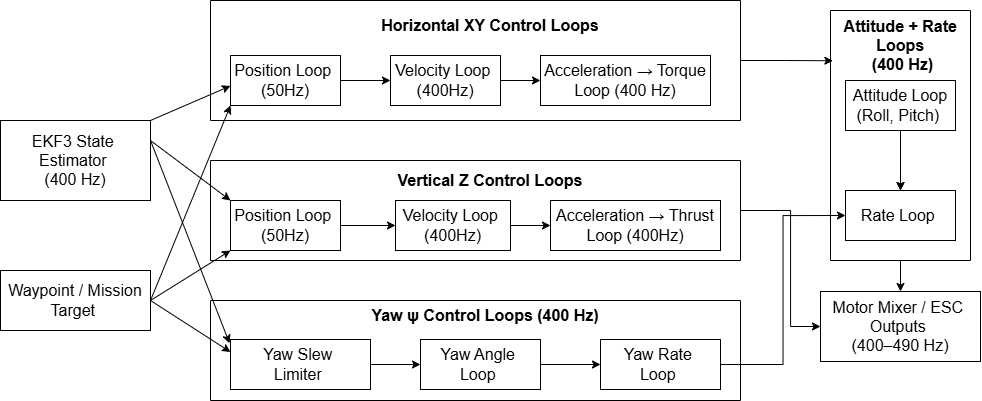}
    \caption{\ac{RE}'d ArduCopter flow.}
    \label{fig:arducopter-flow}
\end{figure}

The ArduCopter flow is illustrated in Figure~\ref{fig:arducopter-flow}, where state estimator and target information are sent for horizontal, vertical, and yaw control loops, which then are used for attitude and rate calculation for the motor/\ac{ESC} control. 
Sensor data from the \ac{IMU}, barometer, magnetometer, and GPS are fused using the \ac{EKF} to estimate the vehicle state at time $t$ in the NED frame, which is described in terms of position ($p$), velocity($v$), and acceleration ($a$) in the three dimensions $(x,y,z)$ as depicted by Eq.~\ref{eq:state}.

\begin{align}\label{eq:state}
p(t) = 
\begin{bmatrix}
x \\ y \\ z
\end{bmatrix} (cm), \quad
v(t) = 
\begin{bmatrix}
v_x \\ v_y \\ v_z
\end{bmatrix} (cm/s), \quad
a(t) = 
\begin{bmatrix}
a_x \\ a_y \\ a_z
\end{bmatrix} (cm/s^2).
\end{align}

The filtered state estimates, at each time iteration $t$, are combined with the user commands received via MAVLink to compute the distance between the target and current states formulated as errors for the control models, as depicted by Eq.~\ref{eq:error}. The states are subsequently updated using the computed errors through four nested loops for position, velocity, attitude, and rate control. 

\begin{align}\label{eq:error}
e(t) = \text{target}(t) - \text{measurement}(t).
\end{align}

Arducopter applies the discrete \acf{PID} controller on the feedback variables as formulated using Eq.~\ref{eq:FFPIDMode}. 

\begin{align}\label{eq:FFPIDMode}
f(t) = k_P e(t) + k_I \sum_{i=0}^{t} e_i \Delta t  + k_D \frac{e_t - e_{t-1}}{\Delta t} + k_{F} \cdot \text{target}(t).
\end{align}
where $k_P$,$k_I$, and $k_d$ are constants used for each state variable. 
Next, the feed-forward is used for guided and auto modes for planned path and mission control tasks,. 

ArduCopter employs a cascaded control architecture with hierarchically scheduled loops for smooth transitions and stability, operating at distinct frequencies. The target states $(p_{t_x}, p_{t_y}, p_{t_z})$ from mission are processed through successive controllers: Position (50~Hz) $\rightarrow$ Velocity (100~Hz) $\rightarrow$ Attitude (400~Hz) $\rightarrow$ Rate (400~Hz) $\rightarrow$ Motor Outputs (400–490~Hz). 
Thus, \textit{desired states} $(p_{d_x}, p_{d_y}, p_{d_z})$ are formed iteratively from high-level targets, while \textit{controller outputs} $(p_{c_x}, p_{c_y}, p_{c_z})$ feed inner loops or the motor mixer. The ArduPilot scheduler executes slow (position, EKF) and fast (attitude, rate) tasks deterministically to maintain timing consistency. 

\subsection{Cascaded Horizontal Control} 
The horizontal motion controller operates as a 2D cascaded loop. The \textbf{outer position controller} computes the velocity setpoints from position errors as in Eq.~\ref{eq:positioncontrol}. 

\begin{align}
p_{dxy}(t) = p_{txy}(t) - p_{offxy}(t), 
\end{align}
\begin{align}
e_{pxy}(t) = p_{txy}(t) - p_{cxy}(t), 
\end{align}

\begin{align}\label{eq:positioncontrol}
v_{dxy}(t) &= k_{P,xy}\,e_{pxy}(t)
   + k_{I,xy}\!\sum_{i=0}^{t} e_{pxy}(i)\,\Delta t \notag \\[2pt]
   &\quad + k_{D,xy}\,\frac{e_{pxy}(t)-e_{pxy}(t-1)}{\Delta t}.
\end{align}

\noindent
where $p_{txy}(t)$, $p_{cxy}(t)$, and $p_{dxy}(t)$  are the target,current and desired horizontal positions at time $t$, respectively. The term \(\mathbf{p}_{off,xy}\) act as small corrections that keep motion smooth when the EKF reference frame changes. These offsets match the correction terms and help maintain stable control during estimator updates.


The resulting $v_{dxy}(t)$ serves as input to the \textbf{inner velocity controller}, which generates the acceleration target as follows. 
\begin{align}
v_{dxy}(t) &= v_{txy}(t) - v_{offxy}(t)
\end{align}
\begin{align}
e_{vxy}(t) &= v_{txy}(t) - v_{cxy}(t), \\[4pt]
a_{dxy}(t) &= k_{Pv,xy}\,e_{vxy}(t)
   + k_{Iv,xy}\!\sum_{i=0}^{t} e_{vxy}(i)\,\Delta t \notag \\[2pt]
   &\quad + k_{Dv,xy}\,\frac{e_{vxy}(t)-e_{vxy}(t-1)}{\Delta t}.
   \label{eq:horizentalacceleration}
\end{align}

where $v_{txy}(t)$, $v_{cxy}(t)$ and $v_{dxy}(t)$ are the target, current and desired horizontal velocity, and ($k_{Pvxy}, k_{Ivxy}, k_{Dvxy}$) are the Kalman gains for the horizontal velocity. Feed-forward and offset terms are added to the velocity and acceleration targets to anticipate autopilot demands and to compensate for small estimation biases, improving controller smoothness and tracking accuracy.

\begin{align}
\mathbf{a}_{txy}(t)
&= \mathbf{a}_{dxy}(t)
 + \mathbf{a}_{offxy}(t), \
\end{align}
where $\mathbf{a}_{dxy}(t)$ is the desired acceleration computed from the velocity control loop, 
and $\mathbf{a}_{offxy}(t)$ represents a correction offset added to maintain smooth continuity across EKF resets, frame shifts, or mode transitions. 

Acceleration commands are then constrained by configured limits as in Eq.~\ref{eq:horizentalaccelerationconstraint} below:
\begin{equation}\label{eq:horizentalaccelerationconstraint}
a_{txy}(t) = \text{constrain}\!\big(a_{txy}(t),\, -a_{\max,xy},\, a_{\max,xy}\big),
\end{equation}
ensuring bounded responses ($[-a_{\max,xy}, a_{\max,xy}]$) under aggressive maneuvers. 
The final target acceleration in the horizontal plane is obtained by summing the instantaneous target, desired, and offset accelerations as implemented in the above control equation.

The resulting $\mathbf{a}_{txy}(t)$ serves as the effective acceleration command for attitude generation and is later constrained by the maximum lean angle limit.

\subsection{Cascaded Altitude Control} 

Altitude control is performed through three cascaded PID loops in the downward-positive \ac{NED} frame. The outer position controller computes the desired climb rate:

\begin{align}
p_{dz}(t) &= p_{tz}(t) - \left[p_{offz}(t) + p_{terrain}(t)\right], \
\end{align}
\begin{align}
    e_{pz}(t) &= p_{tz}(t) - p_{cz}(t),\
\end{align}
\begin{align}
v_{dz}(t) &= k_{P,z}\,e_{pz}(t)
         + k_{I,z}\!\sum_{i=0}^{t} e_{pz}(i)\,\Delta t
         + k_{D,z}\,\frac{e_{pz}(t)-e_{pz}(t-1)}{\Delta t}.
\end{align}
where $p_{tz}(t)$,$p_{cz}(t)$,$p_{dz}(t)$ are the target, current and desired vertical position at time $t$ respectively, and and $(k_{P,z}, k_{I,z}, k_{D,z})$ are the Kalman gains constants for the vertical position, where epz is the altitude error. 

\begin{align}
v_{d_z}(t) &= v_{t_z}(t) - \left[v_{off_z}(t) + v_{terrain}(t)\right], 
\end{align}
The velocity controller converts this into acceleration:
\begin{align}
e_{v,z}(t) &= v_{tz}(t) - v_{cz}(t), \
\end{align}
\begin{align}
a_{dz}(t) &= k_{Pv,z}\,e_{v,z}(t)
         + k_{Iv,z}\!\sum_{i=0}^{t} e_{v,z}(i)\,\Delta t
         + k_{Dv,z}\,\frac{e_{v,z}(t)-e_{v,z}(t-1)]}{\Delta t}.
\end{align}
where $v_{tz}(t)$, $v_{cz}(t)$ and $v_{dz}(t)$ are the target, current and desired vertical velocity at time $t$ respectively, and $(k_{Pvz}, k_{Ivz}, k_{Dvz})$ are the Kalman gains constants for the vertical velocity, and the acceleration is refined with feed-forward, offset, and terrain-compensation terms to correct for estimation bias and elevation variations. This results in smooth and robust altitude tracking under varying terrain or sensor drift. 
Finally, the acceleration command is mapped to thrust using Eq.~\ref{eq:computethrotle} below.
\begin{align}
e_{a,z}(t) &= a_{tz}(t) - a_{cz}(t), \
\end{align}
\begin{align}\label{eq:computethrotle}
T_{\mathrm{in}}(t)&= k_{Pa,z}\, e_{a,z}(t)
     + k_{Ia,z}\!\sum_{i=0}^{t} e_{a,z}(i)\,\Delta t 
     + k_{Da,z}\,\frac{[e_{a,z}(t)-e_{a,z}(t-1)]}{\Delta t}.
\end{align}

\begin{align}\label{eq:accelerationatz}
a_{t_z}(t) &+= a_{d_z}(t) + a_{off_z}(t) + a_{terrain}(t).
\end{align}
where $a_{tz}(t)$ represents the target vertical acceleration, and $a_{cz}(t)$ denotes the current measured vertical acceleration at time $t$. Tuple $(k_{Paz}, k_{Iaz}, k_{Daz})$ correspond to the gain terms regulating vertical acceleration. 
 $p_{terrain}(t)$, $v_{terrain}(t)$, and $a_{terrain}(t)$ are derived from the terrain altitude estimation. Additional offset components $(p_{off_z}, v_{off_z}, a_{off_z})$ are included to correct EKF drifts or frame reference shifts. The compensated vertical target acceleration $a_{tz}(t)$ is then used as the input to the acceleration PID controller, producing a normalized thrust command $T_{in}(t) \in [0,1]$.

The controller operates in the thrust domain, whereas the \acp{ESC} and motors function in the throttle domain (PWM duty cycle). Thus, the normalized thrust command $T_{in}(t)$ is adjusted with a steady-state hover term $T_{\text{hover}}$ and further compensated for gravity and tilt effects to yield the throttle command $T_{out}(t)$, as given in Eq.~\ref{eq:thrustout}:
\begin{equation}\label{eq:thrustout}
T_{out}(t) = \frac{T_{\text{hover}} + T_{in}(t)}{\cos \alpha(t)},
\end{equation}
where $\alpha(t)$ is the thrust–vector tilt angle. Here, $T_{\text{hover}}$ represents the nominal throttle required to maintain level hover. $T_{out}(t)$ denotes the final throttle command delivered to the motor mixer and is bounded within safety limits to prevent actuator saturation during aggressive maneuvers.

\subsection{Attitude and Rate controller}

From ~\ref{eq:horizentalaccelerationconstraint} and ~\ref{eq:accelerationatz} to achieve the commanded lateral acceleration, the desired roll and pitch angles are computed from the thrust–vector relationship as follows:
\begin{equation}
\mathbf{t}_d(t) \propto 
\begin{bmatrix}
-\,a_{t,x}(t) \\[2pt]
-\,a_{t,y}(t) \\[2pt]
g - a_{t,z}(t)
\end{bmatrix}.
\label{eq:thrust_vector}
\end{equation}
For small angles, the desired roll and pitch is approximated as:
\begin{equation}
\text{Roll: } \phi(t) \approx \frac{a_{t,x}(t)}{g(t)}, 
\text{Pitch: } \theta(t) \approx -\,\frac{a_{t,y}(t)}{g(t)}
\label{eq:small_angle}
\end{equation}


\noindent \paragraph{Yaw Control and Rate Limiting.} To prevent instantaneous heading jumps, \textit{ArduCopter} applies a first-order slew limiter on the yaw target. The limiter constrains both the instantaneous yaw rate and the incremental change in yaw target per control cycle.

\begin{align}\label{eq:yawclip}
\dot{\psi}_t(t) &= \mathrm{clip}\big(\dot{\psi}_{\mathrm{cmd}}(t), -\dot{\psi}_{\max}, \dot{\psi}_{\max}\big), 
\end{align}
\begin{align}\label{eq:yawupdate}
\psi_t(t) &= \psi_t(t-\Delta t) + \mathrm{clip}\big(\psi_{\mathrm{cmd}}(t) - \psi_t(t-\Delta t), -\Delta\psi_{\max}, \Delta\psi_{\max}\big), \
\end{align}
where $\dot{\psi}_t(t)$ is the limited yaw rate target, and $\psi_t(t)$ is the updated yaw target after applying the rate and slew constraints.\(\psi_{cmd}(t)\) represents the the desired yaw direction that the controller should track.Eq.~\ref{eq:yawclip} controls how fast the yaw can change, while Eq.~\ref{eq:yawupdate} determines how much the yaw angle updates in each control cycle, ensuring smooth and stable heading transitions.
 Typical parameter values are:

\begin{align}
\dot{\psi}_{\max} &= \mathrm{RATE\_Y\_MAX}, \\
\Delta\psi_{\max} &= \mathrm{SLEW\_YAW}\,\Delta t. 
\end{align}

\noindent\paragraph{Yaw Transition Smoothing.} This mechanism ensures smooth heading transitions and prevents aggressive yaw accelerations that could destabilize the attitude controller. where \texttt{RATE\_Y\_MAX} limits how fast the drone can actually spin and \texttt{SLEW\_YAW} limits how quickly the target yaw moves, to avoid jerky setpoint jumps.


\begin{equation}
|\dot{\psi}(t)| \le \min(\mathrm{RATE}_{Y\_MAX}, \mathrm{SLEW}_{YAW}). 
\end{equation}

Incremental yaw updates are then limited by the maximum allowable step $\Delta\psi_{\max}$ per control cycle. 

Input shaping enforces bounded accelerations:
\begin{equation}
\boldsymbol{\omega}_t(t) = \mathrm{clip}\big(\boldsymbol{\omega}_t(t-1) - a_{\max}\Delta t,\ \boldsymbol{\omega}_t(t-1) + a_{\max}\Delta t\big), 
\end{equation}
followed by rate clamping:
\begin{align}
|\omega_{x,y}(t)| &\le \mathrm{RATE}_{RP\_MAX}, \\
|\omega_z(t)| &\le \mathrm{RATE}_{Y\_MAX}. \
\end{align}

Quaternion $q = [q_w, q_x, q_y, q_z]$ is used to represent the 3D orientation, and is computed from the triplet $(\phi, \theta, \psi)$ using a set of trigonometric transformation~\cite{Kuipers1999}. The attitude controller computes the orientation error ($e_{ang}$) between the targeted quaternion $\mathbf{q}_t(t)$ and the measured quaternion $\mathbf{q}_b(t)$ using Eq.~\ref{eq:quaterniontoEuler}.

\begin{equation}\label{eq:quaterniontoEuler}
\mathbf{e}_{\mathrm{ang}}(t) = \mathrm{axisangle}\!\big(\mathbf{q}_t(t) \otimes \mathbf{q}_b^{-1}(t)\big),
\end{equation}
yielding the vector
\begin{align}
\mathbf{e}_{\mathrm{ang}}(t) =
\begin{bmatrix}
e_{\phi}(t) \\[2pt]
e_{\theta}(t) \\[2pt]
e_{\psi}(t)
\end{bmatrix}.
\end{align}

The proportional attitude gains map this to the target angular rate using Eq.~\ref{eq:Eulertorate}:
\begin{equation}\label{eq:Eulertorate}
\boldsymbol{\omega}_t(t) = k_{P,\mathrm{ang}} \odot \mathbf{e}_{\mathrm{ang}}(t).
\end{equation}

An optional square–root controller is implemented to scale large angular errors nonlinearly, ensuring smooth transitions as in Eq.~\ref{eq:smootheulerrate}.

\begin{equation}\label{eq:smootheulerrate}
\boldsymbol{\omega}_t(t) = k_{P,\mathrm{ang}} \odot \mathbf{e}_{\mathrm{ang}}(t)
\sqrt{\frac{\omega_{\max}}{|\mathbf{e}_{\mathrm{ang}}(t)| + \epsilon}}.
\end{equation}
where $w_{max}$ and $\epsilon$ are smoothing constants. 

This mechanism is activated when the attitude error magnitude exceeds a threshold, typically during aggressive maneuvers or recovery from large disturbances. 
Instead of applying a purely linear proportional response, the controller compresses large errors using a square–root scaling function, thereby limiting the angular rate command and preventing actuator saturation or oscillatory behavior during rapid attitude changes.
The inner rate PID loop regulates the actual angular velocities measured by the gyroscope. The rate error is defined as:
\begin{equation}
\mathbf{e}_{\omega}(t) = \boldsymbol{\omega}_t(t) - \boldsymbol{\omega}_c(t),
\end{equation}
where $\boldsymbol{\omega}_t(t)$ represents the target body rates from the attitude controller, and $\boldsymbol{\omega}_c(t)$ represents the measured angular rate by the gyroscope.

The body torque command is then computed using a feed-forward PID-based controller using Eq.~\ref{eq:torqueratecontroller}. 
\begin{multline}\label{eq:torqueratecontroller}
\boldsymbol{\tau}(t) = 
k_{P,\omega}\,\mathbf{e}_{\omega}(t)
+ k_{I,\omega}\!\sum_{i=0}^{t}\mathbf{e}_{\omega}(i)\,\Delta t
+ k_{D,\omega}\,\frac{\mathbf{e}_{\omega}(t) - \mathbf{e}_{\omega}(t-1)}{\Delta t} \\
+ k_{FF,\omega}\,\boldsymbol{\omega}_t(t).
\end{multline}

The proportional, integral, and derivative gains $(k_{P,\omega}, k_{I,\omega}, k_{D,\omega})$ define the primary control response, 
while the feedforward term $k_{FF,\omega}$ enhances responsiveness to rapid command changes. The derivative term provides damping to suppress oscillations and ensure smooth torque actuation.

\subsection{Frame Mixer and Actuator Control Mapping}
The frame mixer converts the collective thrust and torque into motor-level commands:
\begin{multline}\label{eq:motormixer}
\mathbf{u}_m(t) = [\,1 \;\; T_{\mathrm{out}}(t)\,] + M\,\boldsymbol{\tau}(t), 
\qquad 
\mathbf{u}_m(t) \in [0,1]^N,
\end{multline}
where $\mathbf{M}$ is the geometry-dependent((based on drone shape, X or + configuration)) mixing matrix mapping body torques to motor thrusts, $\boldsymbol{\tau}(t) = [\tau_\phi(t),\, \tau_\theta(t),\, \tau_\psi(t)]^\top$, and $\mathbf{1} = [1,\,1,\,\dots,\,1]^\top$ represents the collective-thrust contribution to all motors.

The output shaping stage applies:
battery-voltage compensation, exponential throttle shaping, and minimum/maximum thrust saturation.
The final \ac{PWM} or DShot commands are transmitted to the \acp{ESC}, 
where \ac{PWM} provides analog signal control, and DShot is a digital serial protocol offering higher precision and noise immunity.

\subsection{Control Flow and Scheduling}

The control architecture executes in a pipelined manner, with estimation through the \ac{EKF} operating at 100-400~Hz, attitude and rate control loops at 400-800~Hz, velocity control at 50-100~Hz, and position or mode logic at 10-50~Hz (thus $t$ in the controller takes values $1/10$ to $1/50$). The cooperative scheduler ensures deterministic timing, maintaining consistent loop execution rates with minimal jitter between estimation and control layers. This real-time synchronization enables reliable sensor fusion, responsive actuator commands, and stable vehicle dynamics across diverse flight modes.

Figure~\ref{fig:arducopter-flow} shows the sequence of computation of the control models. Horizontally, the cascaded control follows the sequence: 
Position~$(x,y)$~$\rightarrow$~Velocity~$(x,y)$~$\rightarrow$~Acceleration~$(x,y)$~$\rightarrow$~Lean~Angles~$(\phi,\theta)$~$\rightarrow$~Rate~$(\phi,\theta)$~$\rightarrow$~Torque~$(\tau)$~$\rightarrow$~Motors. 

Vertically, altitude regulation proceeds through 
Position~$(z)$~$\rightarrow$~Velocity~$(z)$~$\rightarrow$~Acceleration~$(z)$~$\rightarrow$~Thrust~$(T_{\text{in}})$~$\rightarrow$~Motors~$\rightarrow$~Throttle~. 
Yaw control operates independently through 
Desired~Yaw~$\rightarrow$~Yaw~Rate~$\rightarrow$ Torque~$(\tau)$~$\rightarrow$~Motors. 

Each control loop refines the output of its predecessor, producing smooth transitions between layers. 
Integrated shaping filters, safety constraints, and compensation mechanisms further enhance stability, robustness, and responsiveness under dynamic flight conditions.

\section{Model Security Analysis}
\label{sec:analysis}

An attacker can compromise ArduPilot control systems through four forms of manipulation: (1) injection of MAVLink control commands, (2) modification of control parameters, (3) manipulation of the sensors' behaviors, and (4) exploitation of exception-handling mechanisms. In the following, we outline each manipulation type and its potential implications for system resilience. Future work will involve simulation-based and analytical evaluations to characterize the scope and boundaries of these attack vectors.

\subsection{Injection of MAVLink control commands}

Given that the MAVLink protocol provides little to no built-in protection, an adversary within radio range can spoof \ac{GCS} and \ac{RC} to inject control messages that the vehicle will accept as if they originated from a legitimate \ac{GCS}. 
\footnote{MAVLink 1 lacks authentication; in MAVLink 2, signing is optional and often disabled, leaving the vehicle unable to verify message origin or integrity and enabling command-level spoofing.} 
An adversary with this capability can effect stealthy takeovers, for example, by uploading a new mission (\texttt{MISSION\_COUNT}/\texttt{MISSION\_ITEM\_INT} followed by \texttt{MISSION\_START}), by modifying home/return-to-launch behavior via \texttt{SET\_HOME\_POSITION}, or by falsifying pilot input through \texttt{RC\_CHANNELS\_OVERRIDE}. The attacker may also issue \texttt{MAV\_CMD\_DO \_REPOSITION} to redirect the vehicle to chosen coordinates while masking the takeover by forging benign \texttt{HEARTBEAT} traffic. 
This attack vector is possible due to MAVLink's lightweight message-oriented design together with successful manipulation, which requires minimal effort but the knowledge of message identifiers, field formats, and timing, rather than firmware exploits. Thus, it has been widely examined in the literature (e.g., \cite{10740195}). 

\subsection{Manipulating Configuration Parameters}
\label{ssec:param_manipulation}

ArduCopter exposes a wide range of configuration parameters controlling gains, limits, and thresholds. The limits (e.g., maximum acceleration, maximum lean angles, altitude limits) prevent the autopilot from exceeding hardware or safety boundaries, and the thresholds (e.g., EKF variance thresholds via \texttt{FS\_EKF\_THRESH}, failsafe triggers, sensor health indicators) determine when control or estimation algorithms fall back to safe modes. Noise thresholds and filter coefficients directly affect sensor validation and the sensitivity of the state estimator to outliers or spoofed data. Manipulating these values (e.g., raising thresholds so that faults or variances are ignored, or disabling limits for aggressive flight modes) can mask faults, allow attacks to persist longer, or inadvertently destabilize the vehicle. In the following, we describe the main parameters that could be manipulated (Full set of parameters is available at~\cite{Ardupilotparam}). 

\noindent{\bf Update loop frequency.} The control loop frequencies can be adjusted via MAVLink commands, which effectively modify the time-step parameter $t$ in the PID controllers, as defined in Eq.~\ref{eq:FFPIDMode}. For instance, the parameter \texttt{SCHED\_LOOP\_RATE} defines the main control loop frequency in hertz (Hz), typically ranging from 50 to 1000 Hz, with a default value of 400 Hz. Increasing this rate enhances control smoothness and responsiveness but imposes a higher computational load. Excessively high rates can saturate the processor, leading to thread synchronization issues between control computation and physical actuation. Conversely, setting the rate too low may introduce substantial delays between sequential updates, resulting in unstable motion. 

\noindent{\bf Changing the Rate-gain used in the \ac{PID} control.} The control system employs a cascade of \ac{PID} controllers, as illustrated in Eq.~\ref{eq:FFPIDMode}, with three gain parameters: $k\_P$, $k\_I$, and $k\_D$. ArduPilot provides default values for these gains, which can be modified via MAVLink commands. For example, sending a MAVLink \texttt{PARAM\_SET} command with the parameter name \texttt{ATC\_RAT\_PIT\_I} adjusts the pitch-axis rate controller's integral gain $k_I$, as defined in Eq.~\ref{eq:torqueratecontroller}. 
Accordingly, the nominal Rate PID torque is computed by Eq.~\ref{eq:rate_PID_torque}. 
\begin{equation}\label{eq:rate_PID_torque}
\tau[k] = PID[ e_\omega[t]] 
\end{equation}
where $e_\omega[t]$ is the angular rate error at time step $t$. When the controller gains are modified or an additional torque bias is applied, the resulting torque can be calculated by Eq.~\ref{eq:tau_k}. 
\begin{equation}\label{eq:tau_k}
\tau[k] = PID' e_\omega[t] + \tau_{adv}[t]
\end{equation}
where PID' represents the modified PID controller and $\tau_{\mathrm{adv}}$ denotes an additive torque input. Adjusting the gains shifts the closed-loop pole locations, which can render the system underdamped, unstable, or sluggish. The additive term $\tau_{\mathrm{adv}}$ provides a mechanism for injecting steady-state or oscillatory torque, enabling targeted manipulation of the system's dynamic response.

\noindent{\bf Actuator Limit Modification (Saturation and Windup).}
The control loops typically enforce actuator saturation limits to ensure that control commands remain within physically safe bounds. These limits, however, can be intentionally modified to induce undesired or unstable behavior. For instance, the nominal actuator saturation in Eq.~\ref{eq:motormixer} is expressed in Eq.~\ref{eq:u_sat_t}. 

\begin{equation}\label{eq:u_sat_t}
u_{\mathrm{sat}}[t] = \operatorname{sat}\big(u[t],\, u_{\min},\, u_{\max}\big),
\end{equation}
where $u[t]$ denotes the control input, and $u_{\min}$ and $u_{\max}$ represent the lower and upper saturation limits, respectively. Altering these limits either relaxing or tightening them results in an updated actuator constraint given by Eq.~\ref{eq:u_satp_t}. 

\begin{equation}\label{eq:u_satp_t}
u_{\mathrm{sat}}'[t] = \operatorname{sat}\big(u[t],\, u_{\min} + \Delta u_{\min},\, u_{\max} + \Delta u_{\max}\big).
\end{equation}

Such modifications can significantly affect the system's integral windup behavior. Relaxed bounds may permit commands beyond the actuator's safe operating range, while overly restrictive bounds may cause premature saturation, clipping legitimate control efforts, and degrading system performance.

Relaxed limits (\(\Delta u_{\min} < 0\) or \(\Delta u_{\max} > 0\)) let outputs exceed safe bounds, risking actuator damage or instability. 
This behavior reflects standard actuator saturation logic and illustrates how tampering with bounds can degrade UAV stability.

\noindent{\bf Manipulating sensors' behavior.} ArduPilot allows modification of the bounds and thresholds used in its Kalman filter (see Section~\ref{sec:architecture}), which are designed to reject or downweight outliers and noisy measurements. Modifying these parameters directly influences the UAV's estimation and control performance, affecting position, velocity, and actuators. For instance, increasing the value of \texttt{EKF2\_ACC\_NOISE} reduces the EKF's reliance on accelerometer measurements, causing it to place greater trust in predicted state estimates. This can lead to delayed or inaccurate state updates if the model does not accurately reflect the true system dynamics.

\subsection{Manipulating Sensors' Input}

The control system integrates measurements from accelerometers, gyroscopes, magnetometers, barometers, and GPS sensors. The \ac{EKF} is employed to mitigate measurement noise and external disturbances through multi-sensor fusion, enabling accurate and robust state estimation for feedback control. However, these sensors may be compromised by falsified readings or signal suppression induced through electromagnetic or acoustic interference~\cite{electronics13020393}. Such disturbances can propagate through the estimation and control layers, adversely affecting system stability, safety, and reliability. We present, in the following, several examples of sensor reading manipulation.

\noindent{\bf (i) GPS Spoofing.} An attacker can feed falsified sensor GPS data so that the EKF trusts a biased measurement and gradually drifts from the true state. Let the true measurement be $z_{\mathrm{true}}(t)=h(x(t))$, where $h(\cdot)$ maps the state to the ideal observation, and let the spoofed input be $z_{\mathrm{gps}}(t)$ as follows.

\begin{equation}\label{eq:z_gps_t}
z_{\mathrm{gps}}(t) = h(x(t)) + a(t) + \nu(t),
\end{equation}
where $a(t)$ is an attacker-controlled injection (bias, drift, or replay) and $\nu(t)$ is sensor noise. 
The EKF residual becomes
\begin{equation}
\tilde{y}(t) = z_{\mathrm{gps}}(t) - H(t)\hat{x}(t|t^{-}),
\end{equation}
and the attack propagates through the Kalman update as
\begin{equation}
\Delta\hat{x}(t) = K(t)a(t).
\end{equation}
A slowly varying $a(t)$ can remain within detection thresholds while introducing a persistent bias $K(t)a(t)$ in the estimated state.

\noindent{\bf (ii) IMU Spoofing.}
IMU spoofing corrupts the acceleration and angular-rate feedback used in control loops.
Let $a_d(t)$ denote the desired acceleration and $a_c(t)$ the measured acceleration used by the controller.
Under spoofing, measured acceleration becomes 
\begin{equation}
a_c(t) = a_d(t) + a_{\mathrm{sp}}(t) + \nu_a(t),
\end{equation}
where $a_{\mathrm{sp}}(t)$ represents injected interference (e.g., EMI or acoustic) and $\nu_a(t)$ is sensor noise.
The resulting error $e_a(t) = a_d(t) - a_c(t)$ drives a biased thrust command, leading to incorrect motor responses.
Similarly, for gyroscope spoofing,
\begin{equation}
\omega_c(t) = \omega_d(t) + \omega_{\mathrm{sp}}(t) + \nu_\omega(t),
\end{equation}
which causes attitude drift and instability in the rate loop.

\noindent{\bf (iii) Forged GPS Streams.}
If $p_{dxy}(t)$ and $p_{dz}(t)$ represent the desired position and altitude, the spoofed GPS readings become
\begin{equation}
p_{\mathrm{gps}}(t) = p_{dxy}(t) + b_p(t) + \nu_p(t),
\end{equation}
\begin{equation}
h_{\mathrm{gps}}(t) = p_{dz}(t) + b_h(t) + \nu_h(t),
\end{equation}
where $b_p(t), b_h(t)$ are attacker biases and $\nu_p(t), \nu_h(t)$ are measurement noise.
These can corrupt EKF residuals and yield erroneous correction terms $K(t)[b_p(t), b_h(t)]^\top$, leading to false navigation setpoints.

\subsection{Exception Handling}

ArduPilot implements a set of conditions to detect faulty states and trigger appropriate failsafe actions, such as switching flight modes, increasing throttle, initiating an emergency landing, or disarming the vehicle. We discuss in the following some of these exceptions.

\noindent{\bf Crash reporting.} The system declares a crash condition when all of the following criteria are met: (a) the vehicle is armed and crash detection is enabled, (b) it is not in standby or forced-flight mode, (c) it is operating in an angle-stabilized flight mode or during a flip, (d) it is not in autorotation mode, (e) the attitude indicates a lean angle of at least $15^\circ$, (f) the acceleration magnitude is below $3~\text{m/s}^2$, (g) the angular error between the desired and actual thrust vectors exceeds $30^\circ$, and (h) the horizontal velocity is less than $10~\text{m/s}$. When these conditions are satisfied, the system increments the crash counter and evaluates it against a given threshold, which ensures that the criteria persists for at least two seconds before a crash is confirmed and the corresponding failsafe response is triggered.

\noindent{\bf Failsafe Check.} The exception handler monitors the control loop to detect CPU stalls or main-loop lockups. A fault is declared when the (a) failsafe mechanism is enabled, (b) the system is not already in failsafe, (c) the motors are active, and (d)  more than two seconds have elapsed since the last loop execution. Upon detection, the event is logged, and the mitigation procedure begins by reducing motor outputs to minimum thrust and disarming the vehicle. If the system remains in failsafe and the condition persists for an additional second, a full motor shutdown is enforced to ensure safe termination of operation.


\noindent{\bf Manipulation of the exception handling.} ArduPilot employs a classical exception-handling approach, in which an exception is triggered only when a specific set of conditions is satisfied. This contrasts with the traditional failsafe principles, which state that there is a fault unless certain conditions are met. In ArduPilot, these conditions appear to have been determined empirically. More rigorous simulation studies could potentially identify scenarios, such as copter crashes, that occur even when the defined conditions are not fully satisfied.

\section{Conclusions}\label{sec:conclusions}
This work deconstructed the internal control architecture of ArduCopter, revealing a hierarchically cascaded PID framework governed by deterministic scheduling. Through reverse engineering, we demonstrated that seemingly legitimate configuration parameters, sensor feeds, and MAVLink inputs can be exploited to destabilize flight behavior without altering the firmware. By tracing spoofed sensor data and injected commands through the \ac{EKF} and cascaded control loops, we showed how the control stack can be coerced into producing unsafe yet operationally valid actuator outputs. These insights highlight that UAV resilience must extend beyond communication-layer security to include estimator and control-aware safeguards within the autopilot core, ultimately bridging control theory, embedded assurance, and cyber defense to preserve flight integrity under adversarial conditions.

\bibliographystyle{IEEEtran}
\bibliography{Bibliography}

\end{document}